\documentclass[journal]{IEEEtran}
\ifCLASSINFOpdf
  \usepackage[pdftex]{graphicx}
\else
\fi
\usepackage[cmex10]{amsmath}
\usepackage{amssymb}
\usepackage{amsmath}
\usepackage{tabularx}
\newcolumntype{C}{>{\centering}X}
\usepackage{filecontents,lipsum}
\usepackage[noadjust]{cite}
\usepackage{graphicx}
\usepackage{subfig}
\begin{document}
%
\title{Secured Outsourced Content Based Image Retrieval Based on Encrypted Signatures Extracted From Homomorphically Encrypted Images}
%
%
%

\author{Reda~Bellafqira,
        Gouenou~Coatrieux,
        Dalel~Bouslimi,
        Gw\'enol\'e~Quellec
        and~Michel~Cozic
\thanks{R. Bellafqira, G. Coatrieux and  D. Bouslimi are with the Institut Mines-Telecom, Telecom Bretagne, Unité INSERM 1101 Latim, 29238 Brest Cedex, France (e-mail: reda.bellafqira@telecom-bretagne.eu; gouenou.coatrieux @telecom-bretagne.eu; dalel.bouslimi@telecom-bretagne.eu).}
\thanks{G. Quellec is with Inserm, UMR 1101, F-29200 Brest, France (e-mail: gwenole.quellec@inserm.fr).}
\thanks{M. Cozic is with MED.e.COM, Plougastel Daoulas 29470, France (e-mail: mcozic@wanadoo.fr).}
\thanks{This work was supported in part by LabCom and Region Bretagne}
}

\maketitle

\begin{abstract}
In this paper, we present a novel Secured Outsourced Content Based Image Retrieval solution, which allows looking for similar images stored into the cloud in a homomorphically encrypted form. Its originality is fourfold. In a first time, it extracts from a Paillier encrypted image a wavelet based global image signature. In second, this signature is extracted in an encrypted form and gives no clues about the image content. In a third time, its calculation does not require the cloud to communicate with a trusted third party as usually proposed by other existing schemes. More clearly, all computations required in order to look for similar images are conducted by the cloud only with no extra-communications. To make possible such a computation, we propose a new fast way to compare encrypted data, these ones being encrypted by the same public key or not, and using a recursive relationship in-between Paillier random values when computing the different resolution levels of the image wavelet transform. Experiments conducted in two distinct frameworks: medical image retrieval with as purpose diagnosis aid support, and face recognition for user authentication; indicate that the proposed SOCBIR does not change image retrieval performance. 
\end{abstract}


%
\IEEEpeerreviewmaketitle

\section{Introduction}
\label{intro}
Cloud computing has emerged as a successful paradigm allowing individuals and companies to flexibly store and process large amounts of data without a need to purchase and maintain their own networks and computer systems. In healthcare for example, different  initiatives aim at sharing medical images and Personal Health Records (PHR) in between health professionals or hospitals with the help of the cloud \cite{marcuhealthcare}\cite{silvasharing}\cite{vincent2016privacy}. In order, to take advantage of such volume of data and of the medical knowledge it represents, several Content Based Image Retrieval (CBIR) methods have been introduced to support decision making \cite{quellec2010wavelet}. In a general way, the basic concept of a CBIR system stands in the retrieval of images into a database that are similar to a request image. In medical imaging, knowing the diagnosis associated to these similar images can give clues to the physician before he  accesses or analyses the image. Another application that can take advantage of CBIR corresponds to user authentication based on face recognition, where a user is identified through the comparison of his biometric data with those stored and managed in a database into the cloud by a third party \cite{halderman2009lest}. If there is a match, the user can log into his account or obtain the service provided by the third party.

Beyond their innovative character, such cloud applications must take into account data security issues that are increased especially in terms of confidentiality and privacy. Indeed, data outsourcing involves at the same time that users loss the control on their own data \cite{chow2009controlling}. Recent news show clear evidence that outsourced storage is not safe against privacy threats, these ones being external  (e.g., hackers \cite{WinNT}) or internal \cite{halderman2009lest}\cite{lewis2014icloud}. There is thus an interest to develop Secured Outsourced Content Based Image Retrieval (SOCBIR) methods.

Different approaches have been proposed to secure CBIR methods. They can be differentiated depending on if they work on local or global image signature, and on the way they make interacting encryption with image signature computation and comparison. Regarding the way signatures are handled, one first solution consists in outsourcing an image along with its signature  calculated \textit{a priori} \cite{lu2009enabling}. In this scenario, the server (or equivalently the cloud) only conducts signatures' comparison. Images are encrypted with algorithms such as AES or 3-DES and cannot be manipulated unless they are decrypted. The CBIR process is thus shared between the client and the server. An alternative to this approach proposes to extract signatures or features directly from encrypted images. To do so, most solutions make use of homomorphic encryption the interest of which is that it allows performing operations (e.g. $"+"$, $"\times"$) onto encrypted data with the guarantee that the decrypted result equals the one carried out onto unencrypted data. To the best of our knowledge, Erkin \textit{et al}. \cite{erkin2009privacy} proposed the first privacy-preserving biometric face recognition scheme securing the Eigenfaces recognition algorithm. However, this scheme only ensures the confidentiality of the query image due to the fact the server compares the signature extracted from the encrypted image it received to the ones of images stored in its database under an un-encrypted form. In \cite{hsu2012image}, SIFT features are extracted from an image encrypted with the Paillier homomorphic cryptosystem. This method needs 1-round of communication between the server and the user for features' extraction and much more to compare image signatures. This scheme has also been proved unfeasible in terms of computational complexity with some security weaknesses \cite{schneider2014notes}. The authors of \cite{schneider2014notes} improve this scheme with the help of fully homomorphic encryption and secure multiparty computation but with as consequence a great increase of the storage, computation and communication complexities. Another improvement suggested in \cite{qin2014towards}, stands on the use of order-preserving encryption (OPE \cite{boldyreva2009order})  and three non-colluding clouds. It should be noticed that both solutions given in \cite{schneider2014notes} and \cite{qin2014towards} do not well preserve the performance of the original SIFT features. To overcome this issue, \cite{qin2016secsift} and \cite{wang2016catch} propose a secure SIFT feature set based on two independent cloud servers. If the complexity of communication with the user is reduced, it assumes that the two servers do not collude. In \cite{bai2014surf}, the interest is given to SURF features, another kind of local image features, and to their extraction from homomorphically encrypted images. As the previous schemes, this solution requires highly interactive communications between the server and the user in order to extract the image signature by executing certain basic operations in the encrypted domain the server cannot conduct alone without accessing to the data in clear (e.g., division, square root). \cite{wang2016secure} proposed to reduce the communication complexity of \cite{bai2014surf} with the help of somewhat homomorphic encryption (SHE) and two non-colluding cloud servers. But using SHE also limits the number of both addition and multiplication operations one can execute repeatedly over cipher-texts (an issue linked to the SHE depth - see \cite{brakerski2011fully} for more details). \cite{bellafqira2015content} \cite{bellafqira2016end} were the first to introduce a global image signature based SOCBIR. Rather than extracting some local characteristics as SIFT and SURF features, these schemes sums up an image to the distribution of its wavelet coefficients in different sub-bands. Such a signature has been shown more optimal in terms of retrieval performance when the image texture plays a major role, which is often the case in medical imaging.

\begin{figure}[t]
\begin{center}
\centering
  \subfloat[]{\label{fig:edge-a}\includegraphics[scale=0.25]{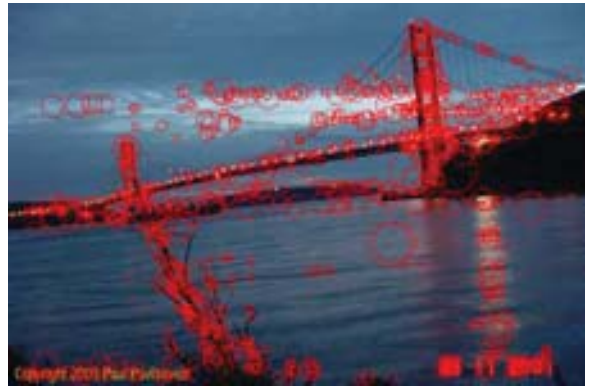}}
  \hspace{5pt}
  \subfloat[]{\label{fig:edge-a}\includegraphics[scale=0.25]{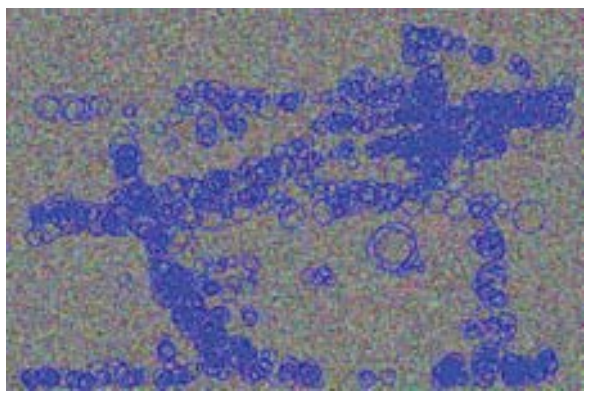}}
  \hspace{5pt}
  \subfloat[]{\label{fig:contour-b}\includegraphics[scale=0.4]{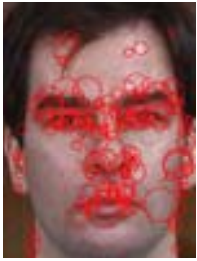}}
  \hspace{5pt}
  \subfloat[]{\label{fig:contour-c}\includegraphics[scale=0.4]{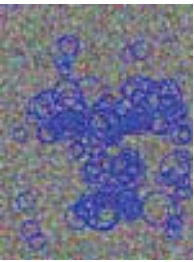}}
\end{center}
\caption{\label{fig1} Illustrative examples issued from \cite{hsu2012image} of SIFT features detection in the plain-text domain (a), (c), and  in the Cipher-text domain (b), (d).}
\end{figure}

Beyond the fact that these SOCBIR schemes require communications in-between the user and one or two servers, they compute image signatures in an un-encrypted form. Even if the image content is not available, knowing its SIFT features may give big clues about it. These features are indeed characteristic points of the image. For  instance, their positions can contributes to the recognition of objects or person in an encrypted image. We give in Fig. \ref{fig1} an example of such a situation. There is thus a need to extract image signatures that are themselves encrypted. Such an issue was very recently addressed in \cite{bellafqira2016end} and by next in \cite{qin2016secsift} \cite{wang2016catch}. But, these solutions exploit at least two cloud servers when computing signatures with as consequence the increase of already high communication requirements.

In this paper, we propose a novel global image signature based SOCBIR. Working with the Paillier cryptosystem, it first allows extracting from an encrypted image a signature in an encrypted form, giving thus no clues about the image content. It is based on two original secure signature calculation and comparison algorithms that are fully conducted by the server (or equivalently the cloud) that is to say without communications with a third party (i.e., the user or another server). Both algorithms take advantage of a new way  to rapidly compare Paillier encrypted data with no communications, even if the data are encrypted with different public keys. In order to make possible the computation of a signature  that corresponds to the histograms of different wavelet sub-bands of the image, we show the existence of a recursive relationship in-between the random values of the Paillier cryptosystem at different resolutions levels of the image wavelet transform.

The rest of this paper is organized as follows. In Section \ref{Section2}, we come back on the definition of the homomorphic encryption algorithm of Paillier and show how to use it so as to rapidly compare encrypted data. Section \ref{Section3} presents the proposed Secure Outsourced CBIR scheme, detailing how to extract an encrypted signature from an encrypted image with no communication between the server or a third party (e.g., the user or a second service provider). Its security and complexity are discussed in Section \ref{SectionDISC}. Performance of the proposed solution are given and compared to the original solution in the clear domain in Section \ref{Section4} considering two application frameworks: diagnosis aid support and user authentication based on facial recognition. Section \ref{SectionCCL} concludes this paper.

\section{PROCESSING OF HOMOMORPHICALLY ENCRYPTED DATA WITH THE PAILLIER CRYPTOSYSTEM }\label{Section2}

This section first presents the Paillier cryptosystem before introducing how to rapidly compare data encrypted or not with the same public key.

\subsection{Paillier cryptosystem}\label{ssec:2-1}

We opted for the asymmetric Paillier cryptosystem because of its additive homomorphic property \cite{paillier1999public}. In this work, we use a fast version of it defined as follows. Let $((g,K_p), K_s)$   be the public/private key pair, such as
    \begin{equation}
      K_p = pq \quad and \quad K_s= (p-1)(q-1)
    \label{(1)}
    \end{equation}
where $p$ and $q$ are two large prime integers. $\mathbb{Z}_{K_p}= \{0, 1,..., K_p-1\}$ and  $\mathbb{Z}_{K_p}^*$ denotes integers that have multiplicative inverses modulo $K_p$. We select $g\in\mathbb{Z}_{K_p^2}^*$ such as
    \begin{equation}
    \frac{g^{K_s}-1 \, mod\, K_p^2}{K_p} \in \mathbb{Z}_{K_p}^*
    \label{(2)}
    \end{equation}

The Paillier encryption of a plain-text $m\in\mathbb{Z}_{K_p}$ into the cipher-text $c\in\mathbb{Z}_{K_p^2}^*$ using the public key $K_p$ is given by
\begin{equation}
        c= E[m,r]= g^m r^{K_p} \mod K_p^2
        \label{(3)}
\end{equation}
where $r \in \mathbb{Z}_{K_p}^*$  is a random integer associated to $m$ making the Paillier cryptosystem probabilistic or semantically secure. More clearly, depending on the value of $r$, the encryption of the same plain-text message will yield to different cipher-texts even though the public encryption key is the same. As introduced in \cite{nassar2015paillier}, it is possible to get a fast version of~\eqref{(3)} by fixing $g=1+K_p$ without reducing the algorithm security. By doing so, the encryption of $m$ into $c$ requires only one modular exponentiation and two modular multiplications
\begin{equation}
c=E[m,r]=(1+mK_p)r^{K_p} mod K_p^2
\label{(4)}
\end{equation}
As we will see in Section \ref{ssec:2-2}, this property will be of importance for the fast comparison of Paillier encrypted data.

Based on the assumption $g=1+K_p$, the decryption of $c$ using the private key $K_s$ is such as
\begin{equation}
m=\frac{(c^{K_s}-1).K_s^{-1} \mod K_p^2}{K_p} \mod K_p
\label{(5)}
\end{equation}

If we consider two plain-texts $m_1$ and $m_2$, the additive homomorphic property of the Paillier cryptosystem allows linear operations on encrypted data like addition and multiplication, ensuring that
\begin{equation}
E[m_1,r_1].E[m_2,r_2] = E[m_1+m_2, r_1.r_2]
\label{(6)}
\end{equation}
\begin{equation}
E[m_1,r_1]^{m_2} = E[m_1.m_2, r_1^{m_2}]
\label{(7)}
\end{equation}

\subsection{Comparing Paillier encrypted data}\label{ssec:2-2}

Even though the additive property of the Paillier cryptosystem allows linear operations, it can not conduct more complex operations like data comparison or modulo reduction. These operations are however essential in signal processing.

In our concern and as we will see in Section \ref{Section3}, comparison of encrypted data is a core operation in SOCBIR. Two distinct classes of solutions can be considered. The first one, which is also the main one studied since a long time, regroups methods that are based on a protocol; a set of interactions  between two parties that do not collude. Many secure protocols are known for comparing two integers, the so called millionaires' problem. The first protocol was proposed by Yao \cite{yao1982protocols} in 1982. It corresponds to the well-known garbled circuit and has been improved many times since. One of the most efficient implementation based on homomorphic encryption is described by T. Veuguen \cite{veugen2012improving}. But, this one, as all the others, requires many communications between parties so as to conduct one comparison. The second kind of approach is communication free and has been recently introduced by Hsu \textit{et al}. in \cite{hsu2012image}. The solution we propose is an improvement of this approach. It allows comparing two encrypted data, encrypted or not with the same public key, in a less complex way. It takes into account the fast implementation of the Paillier cryptosystem of Section \ref{ssec:2-1}.

\begin{itemize}
\item \textbf{Comparing data encrypted  with the same public key}
\end{itemize}

Before presenting our solution, let us come back on the proposal of Hsu \textit{et al}. \cite{hsu2012image}.

Let us consider a client-server relationship where the server possesses two cipher-texts $c_1=E[m_1,r_1 ]$ and $c_2=E[m_2,r_2 ]$ encrypted by the client; $m_1$ and $m_2$ are two integer values the server wants to compare knowing only $c_1$ and $c_2$. The basic idea of \cite{hsu2012image} is not to compare  directly  data but to  establish their relationship (smaller, greater or equal) through the quantization of data from their encrypted versions. In a first time, the client sends along with $c_1$ and $c_2$, a set of thresholds $\{T_i\}_{1\le i \le N}$  that quantize the dynamic range of both integer values $m_1$ and $m_2$. In practice, the client chooses an increasing  sequence of randomly distributed thresholds $\{T_i\in \mathbb{Z}_{K_p}\}_{1\le i \le N}$ and sends $\{E(T_i,r_1 )\}_{1\le i\le N}$ and $\{E(T_i,r_2 )\}_{1\le i\le N}$  to the server. To compare data, the server has first to identify the interval to which belongs each message  $m_u$, $u\in \{1,2\}$. Herein, it is considered that $m_u$ belongs to the $k^{th}_u$ interval, if the closest threshold of $m_u$ is $T_{k_u}$.  To identify $k_u$, the server starts the following iterative process:
\begin{enumerate}
\item \textbf{Comparison of $E[m_u, r_u]$ with one threshold $E[T_i, r_u]$}. Based on the Paillier cryptosystem property $E[m_u, r_u].E[a,b]=E[m_u+a,r_u.b]$ (see \eqref{(6)}), it is possible to compute the distance $D(T_i, m_u,)=T_i-m_u$ by iteratively multiplying $E[m_u,r_u]$ with $E[1,1]$ until the product equals $E[T_i, r_u]$. The number of iterations $inc$ gives $D(T_i,m_u)$ (i.e. $E[T_i,r_u]= E[m_u+inc,r_u]$; $D(m_u,T_i) = inc)$
\item \textbf{Closest threshold of $m_u$}. The closest threshold $T_{k_u}$ is naturally given by the minimum distance in-between $E[m_u,r_u]$ and all encrypted thresholds $\{E[T_i,r_u ]\}_{1\le i\le N}$. Based on the fact $E[m_u,r_u ] g^{inc}=E[m_u+inc,r_u]$, this whole process can be summed up  for any message $m_u$, $u\in \{1,2\}$, by
\begin{equation}
(d_u,k_u)=\arg \min_{i, inc} (E[T_i, r_u] - E[m_u,r_u]g^{inc} \mod K_p^2 )
\label{(8)}
\end{equation}
where $d_u$ is the distance between the message $m_u$ and its closest threshold of index $k_u$.
\end{enumerate}

After having computed $(d_1,k_1)$ and $(d_2,k_2)$ for $m_1$ and $m_2$, respectively, it is possible to determine whether $m_1 \le m_2$ or not without decrypting them. This is, due to the fact that the same thresholds $\{T_i\}_{1\le i\le N}$ are used for $m_1$ and $m_2$. Notice that knowing the relative distance between $m_1$ and $m_2$ gives no ideas  about their respective  values. At the same time, the complexity of this approach depends of the number of threshold.

Compared to \cite{hsu2012image} our proposal does not need an iterative procedure as well as sending several thresholds. On the contrary the user will just have to send to the server two encrypted versions of one single threshold $T$, i.e., $E[T,r_1]$ and $E[T,r_2]$. $T$ will be used as "value of reference" into the dynamics of $m_1$ and $m_2$ for comparison. To directly compute the distance $d_u$ between $m_u$ and $T$ based on $E[m_u,r_u]$ and $E[T,r_u]$, we take advantage of the fast Paillier cryptosystem hypothesis, i.e, $g=1+K_p$, as follows
\begin{equation}
\begin{array}{l}
        d_u = D(T, m_u)=D^{e}(E[m_u, r_u], E[T_u, r_u])\\
        \;\;\;\; \;= \frac{E[T,r_u]E[m_u,r_u]^{-1}-1 \mod K_p^2}{K_p} \mod K_p \\
         \;\;\;\; \;= \frac{g^T r_u g^{-m_u}r_u^{-1}-1 \mod K_p^2}{K_p} \mod K_p \\
         \;\;\;\; \;= \frac{g^{T-m_u} -1 \mod K_p^2}{K_p} \mod K_p \\
         \;\;\;\; \;= T- m_u \mod K_p \\
\end{array}
\label{(9)}
\end{equation}
where $D$ and $D^e$ are two functions that allows computing the $L^1$-distance in the clear and encrypted domains, respectively.

As previously, once $d_1$ and $d_2$ are calculated, one can determine if $m_1 \le m_2$ without any communications with the user. Again, knowing the relative distance between $m_1$ and $m_2$ gives no clues about the values of $m_1$ and $m_2$. Compared to the solution of Hsu \textit{et al}. \cite{hsu2012image}, the computation complexity of which is $O(\frac{K_p}{N})$ (for $N$ thresholds, \cite{hsu2012image} needs  $\frac{K_p}{N}$ iterations to find the closest threshold in the worst case), our proposal is $O(1)$ (one threshold with no iteration).

\begin{itemize}
\item \textbf{Comparing data encrypted with two different public keys}
\end{itemize}

It should be noticed that the above two solutions requires that $m_1$ and $m_2$ are encrypted with the same public key $K_p$ and random value $r_u$. To overcome this issue and to be able to compare two messages $m_1$ and $m_2$ encrypted with different public keys $K_{p1}$ and $K_{p2}$ and two random values $r_1$ and $r_2$ that are respectively defined by two users $U_1$ and $U_2$, for example (i.e., $c_1=E[m_1,r_1]$ and $c_2=E[m_2,r_2]$), we propose to exploit a reference value $P$ in the following way. It is asked to $U_1$ and $U_2$ to encrypt $P$ with same parameters as for $m_1$ and $m_2$, and to send the results, i.e., $E[P,r_1]$ and $E[P,r_2]$, to the server. Based on \eqref{(9)}, the server is able to compute the relative distances $d_1=D(P,m_1 )= P-m_1$ and $d_2=D(P,m_2 )=P-m_2$,   from which it can derive $m_1-m_2$ or, equivalently, if $m_1$ is equal or greater than  $m_2$. As we will see in Section \ref{ssec:3-4}, this last strategy will help us to compare images encrypted by two different users.

\begin{figure}[t!]
\includegraphics[scale=.45]{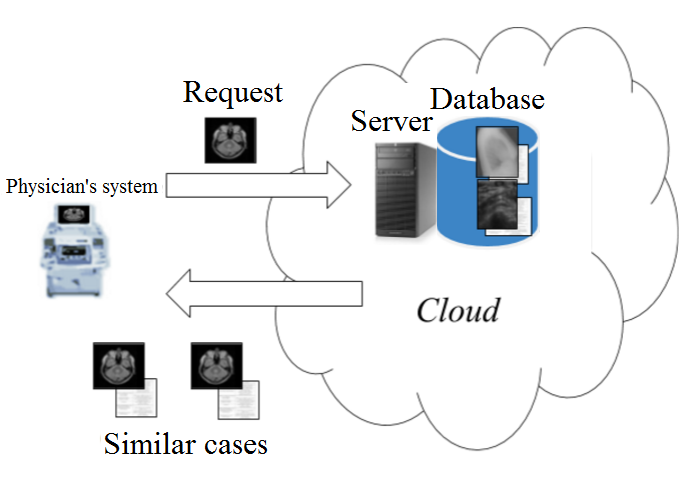}
\centering
\caption{\label{Fig2} Basic interactions of an outsourced content based image retrieval into the cloud.}
\end{figure}

\section{PROPOSED SECURED OUTSOURCED CONTENT BASED IMAGE RETRIEVAL SYSTEM}\label{Section3}

In this section, we first provide a global view of a CBIR system outsourced into the cloud before exposing how to conduct each of its functionalities in a secure manner.

\subsection{From classic CBIR to secure outsourced CBIR}\label{ssec:3-1}

The basic interactions of an outsourced content based image retrieval system are exposed in Fig. \ref{Fig2} in the case of a diagnosis aid support application. Two entities interact: the user (\textit{e.g.} the physician's system) sends a query image to the cloud so as to get an idea about the possible diagnosis. When the cloud receives the request, it extracts or computes from the image a signature it next compares to the ones of the images stored in its database. It then returns to the user the most similar images along with their diagnosis. In this scenario, all computations are conducted in clear by the server. They correspond to the image signature extraction and comparison which are the two basic functionalities of a CBIR system.

\begin{figure}[t]
\includegraphics[scale=.47]{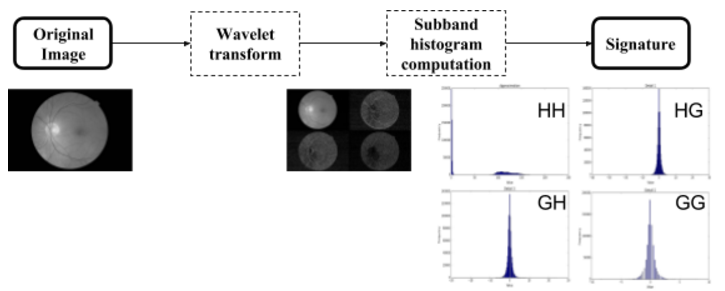}
\caption{\label{Fig3} Image signature computation based on \cite{quellec2010wavelet}. The signature of the image corresponds to the set of its wavelet sub-band histograms up to a given decomposition level. In this example, only one decomposition level is considered. $HH$, $HG$, $GH$ and $GG$ represent the approximation subband and  the horizontal, vertical and diagonal detail subbands, respectively.}
\end{figure}

In this work, we consider the global image signature based CBIR algorithm proposed in \cite{quellec2010wavelet} which sums up an image into a signature that corresponds to the concatenation of the histograms of the different wavelet sub-bands of the image (see Fig. \ref{Fig3}). The computation of this signature relies on two steps:
\begin{itemize}
\item Calculation of the Discrete Wavelet Transform  (DWT) of the image up the $d^{th}$ decomposition level.
\item   Computation of the histograms of the wavelet sub-bands from the level $0$ to the level $d$; and concatenation of the histograms so as to obtain the image signature.
\end{itemize}

The solution we propose and detail in the sequel  has been developed under the following constraints: i) zero interaction in-between the server or any other trusted third parties including  the user when computing and comparing image signatures; and, ii) extract from an encrypted image an encrypted signature. As we will see, our proposal takes advantage of the additive homomorphic property of the Paillier cryptosystem and of the communication free comparison solutions we derived in Section \ref{ssec:2-2}. More clearly, our system allows retrieving similar images with no needs of communications.

\subsection{Wavelet Image Transform in the Paillier Encrypted Domain}\label{ssec:3-2}

The 2D wavelet transform of an image $I$ in the clear domain, considering the wavelet is separable, is defined as \cite{mallat1989theory}
\begin{align}
    a^d(x,y) &=\sum_{l\in \mathbb{Z}}\sum_{l'\in \mathbb{Z}}h(2x-y)h(2y-l')a^{d-1}(l,l')\\
    b^d_u(x,y) &=\sum_{l\in \mathbb{Z}}\sum_{l'\in \mathbb{Z}}w(2x-y)w'(2y-l')a^{d-1}(l,l')
\end{align}
where $a^d(x,y)$ and $b^d_u(x,y)$ are the approximation and detail coefficients of the $d^{th}$ decomposition level, respectively; $(x,y)$ gives the position of the coefficient in the sub-band; $w,w'\in \{h,g \}$ where $h$ and $g$ are the low-pass and high-pass wavelet decomposition filters, respectively; $u\in \{hg,gh,gg\}$ is the index of the detail coefficient sub-bands (i.e., horizontal, vertical and diagonal). In the case $d=0$, $a^0(x,y)$ corresponds to the image pixel $I(x,y)$.

Due to the fact that the Paillier cryptosystem works with plain-text and cipher-text constituted of positive integers in $\mathbb{Z}_{K_p}$ (where $K_p$ is the cryptosystem public key), all image pixels and image processing algorithm's parameters should also be represented by integers. Even if we assume that all pixels are integers, i.e, $I(x,y)\in \mathbb{Z}_{K_p}$, their processing may lead to negative values. In order to represent such values in $\mathbb{Z}_{K_p}$, integer values greater than $\frac{K_p-1}{2}$ will correspond to negative values and the others to positive values.

Similarly, in order to compute DWT in the encrypted domain, the coefficients of the decomposition filters $h$ and $g$, which are real numbers, should be turned into integer values. This can be achieved by scaling and quantizing the coefficients of $h$ and $g$ as follow \cite{zheng2013discrete}
\begin{equation}
    G=[Qg]\, ;\, H =[Qh]
    \label{(12)}
\end{equation}
where $[.]$ is the rounding function and $Q$ is a scaling factor. $Q$ is a parameter defined by the user accordingly to the wavelet transform he wants to exploit.

From this stand point, the encrypted versions of the wavelet approximation and detail coefficients in the encrypted domain are given as (using \eqref{(6)} and \eqref{(7)})
\begin{footnotesize}
\begin{align}
E[A^d(x,y),r^d(x,y)] &=\prod_{l , l'\in \mathbb{Z}}E[A^{d-1}(l,l'),r^{d-1}(l,l')]^{H(2x-y)H(2y-l')}\\
\label{(13)}
E[B^d_u(x,y),r^d_u(x,y)] &=\prod_{l,l'\in \mathbb{Z}}E[A^{d-1}(l,l'),r^{d-1}(l,l')]^{W(2x-y)W'(2y-l')}
\end{align}
\end{footnotesize}
where $A^d(x,y)$ and $B_u^d(x,y)$ are the wavelet approximation and detail coefficients of the $d^{th}$ decomposition level, respectively; $W,W'\in \{H,G\}$ where $H$ and $G$ are the expanded version of the low-pass and high-pass wavelet decomposition filters $h$ and $g$, respectively; $u\in \{HG,GH,GG\}$ indicates the index of the detail coefficient sub-bands (i.e., horizontal, vertical and diagonal). Notice that all multiplications and exponentiation are carried out into  $\mathbb{Z}_{k_p^2}^*$.

\subsection{Calculation of the wavelet sub-band histograms in the encrypted domain}\label{ssec:3-3}

As stated in Section \ref{ssec:3-1}, once the image DWT computed, the next step consists in building the histograms of its different wavelet sub-bands. Let us first recall how to calculate a histogram in the clear domain before explaining how to make the same task in the Paillier encrypted domain so as  to compute a clear histogram and by next an encrypted histogram.

\subsubsection{Histogram computation in the clear domain}\label{sssec:3-3-1}

\begin{figure}[t]
\begin{center}
\includegraphics[scale=.5]{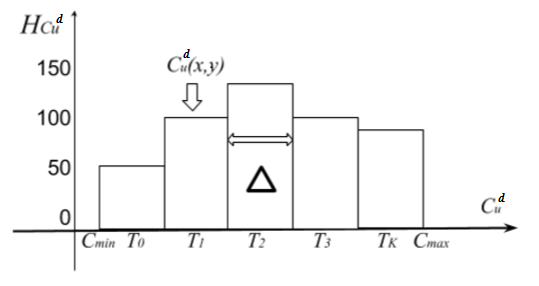}
\end{center}
\caption{\label{Fig4} Histogram $H_{C_u^d}$ of the coefficient sub-band $C_u^d$ in the clear domain, in the case of $K=5$ classes with a coefficient dynamic such as  $C_u^d(x,y)\in [C_{min}, C_{max}]$. $\{T_u\}_{0 \le u \le 4}$ correspond to the class centers.}
\end{figure}

Let us consider $C_u^d (x,y)$ is a wavelet coefficient (either a detail or approximation coefficient) at the position $(x,y)$ in the sub-band $u$, $u\in\{HH,GH,HG,GG\}$,  at the decomposition level $d$. In order to build the histogram $H_{C_u^d}$ of the sub-band $C_u^d$, the coefficient dynamic is first divided into $K$ uniform intervals or classes of width $\Delta$ (see Fig. \ref{Fig4}). By definition, the value $H_{C_u^d} (k)$ indicates the number of coefficients $C_u^d (x,y)$ the values of which belong to the $k^{th}$ interval of $H_{C_u^d}$. More clearly, $H_{C_u^d} (k)$ gives the cardinality of the class $C_k$. If $T_k$ denotes the center of the $k^{th}$ interval of $H_{C_u^d}$, then the class $C_k$ of $C_u^d (x,y)$ is given by:
\begin{equation}
        k = \arg \min_p |C_u^d(x,y)-T_p|
 \label{(15)}
\end{equation}

Another way to construct a histogram is to use a single threshold instead of the K as explained before. The idea behind this approach is that the server uses this threshold as a reference point in order to find the membership classes regardless of the quantification step. To do this, we will consider $T$ as a reference point, then the server computes the distance $d_{x,y}$ between $T$ and each wavelet coefficient $C_u^d(x,y)$, this distance is given as:
\begin{equation}
       		d_{x,y}=T-C_u^d(x,y)
 \label{(15.5)}
\end{equation}			
After the computation of the distances, the objective now is to construct the histogram from the distances between the reference points and all the wavelet coefficients. To do this, we suppose that $d_{x,y}$ and $d_{x',y'}$ present the distances of $C_u^d(x,y)$ and $C_u^d(x',y')$ with $T$ respectively, then it is sufficient to compare $[d_i/\Delta]$ and $[d_j/\Delta]$, where $\Delta$ is the quantization step and $[]$ is the floor operator. By doing so, if $[d_{x,y}/\Delta]$ and $[d_{x',y'} /\Delta]$ are equal then $C_u^d(x,y)$ and $C_u^d(x',y')$  belong to the same class $C_k$ otherwise no. After the comparison between the distances, the server can build an histogram with $K$ uniform classes, where $K$ is given as $K = |C_u^d|/\Delta$ and $|C_u^d|$ presents the cardinal of the dynamic bounds $C_u^d$. 
				
\subsubsection{Clear histogram computation in the encrypted domain}\label{sssec:3-3-2}
The construction of the above histogram in the encrypted domain imposes  to determine the class $C_k$ to which belongs a wavelet coefficient $C_u^d (x,y)$ from its encrypted version $C_u^{de} (x,y)=E[C_u^d (x,y),r_u^d (x,y)]$ where $r_u^d (x,y)$ is a random integer number (see ~\eqref{(3)}). In other words, we need to compute \eqref{(15.5)} in the encrypted domain.

To do so, the user has to send along with the encrypted image the encrypted versions of the histogram class center, i.e., $E[T,r_u^d(x,y)]$. Based on this information and using the encrypted data comparison solution \eqref{(9)} of Section \ref{ssec:2-2}, it is possible to determine the interval or class of a the coefficient $C_u^d (x,y)$ from $C_u^{de}(x,y)$ as follow
\begin{equation}
  d_{x,y}= D^{e}(C_u^{de}(x,y), E[T,r_u^d(x,y)])
  \label{(16)}
\end{equation}
It is important to underline that in order the cloud server computes ~\eqref{(16)} based on \eqref{(9)}, the user  needs to encrypt the threshold $T$ with the appropriate random value $r_u^d (x,y)$ that is to say the same random value associated to the encrypted coefficient $C_u^{de} (x,y)$. It is however possible to find a recursive relation in-between the values  $r_u^d (x,y)$ from the level $0$ to the level $d$ by using the Paillier cryptosystem properties \eqref{(6)} and \eqref{(7)}. Basically, $r_u^d (x,y)$,  $u\in \{HH, GH_d,GH_d,HG_d\}$, is such as
\begin{equation}
    r_u^d(x,y)=\prod_{l,l'\in \mathbb{Z}} r_u^{d-1}(l,l')^{W(2x-l)W'(2x-l')}
    \label{(17)}
\end{equation}
As it can be seen $r_u^d(x,y)$ is independent of the image content. For $d=0$, $r_u^d(x,y)=r(x,y)$, i.e., the random integer value used to encrypt the image pixel $I(x,y)$.

Based on \eqref{(16)} and \eqref{(17)}, the server can build the histogram of one wavelet sub-band ($H_{C_u^d}$). In order to compare two histograms, $H_{C_u^{d(1)}}$ and $H_{C_u^{d(2)}}$ of two images $I^{(1)}$ and $I^{(2)}$, respectively, one just has for instance to apply the $L^1-$distance
    \begin{equation}
    L^1(H_{C_u^{d(1)}}, H_{C_u^{d(2)}})=\sum_{1\le k\le K}|H_{C_u^{d(1)}}(k) - H_{C_u^{d(2)}}(k)|
    \label{(18)}
    \end{equation}
    
\begin{figure}[t]
\begin{center}
\subfloat[]{\label{fig:edge-a}\includegraphics[scale=0.6]{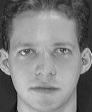}}
  \hspace{5pt}
  \subfloat[]{\label{fig:edge-a}\includegraphics[scale=0.6]{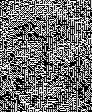}}
  \hspace{5pt}
  \subfloat[]{\label{fig:contour-b}\includegraphics[scale=0.6]{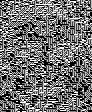}}
  \hspace{5pt}
  \subfloat[]{\label{fig:contour-c}\includegraphics[scale=0.6]{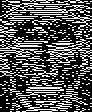}}
\end{center}
\caption{\label{Fig5} Reconstruction attack example of the image (a) from its encrypted wavelet detail coefficients considering two decomposition levels and different quantization steps: (b) $\Delta=32$, (c) $\Delta=64$, (d) $\Delta=128$}
\end{figure}

Even though this solution makes use of encrypted histogram class center  $E[T  ,r]$, it suffers of two security issues. The first one stands on the fact the class cardinalities $\{H_{C_u^d}(k)\}_{1\le k \le K}$ are not encrypted. Along with the awareness the class centers of the histogram are ordered, the histogram $H_{C_u^d}$ appears thus in a clear form. Even if the coefficient dynamic bounds are unknown, if one has some \textit{a priori} knowledge about the signal distribution, he or she will have big clues about the content of the encrypted data. For instance, it is well known that detail wavelet coefficients follow a Gaussian or Laplacian distribution centered onto zero. We give in Fig. ~\ref{Fig5} a reconstruction attack example where the pirate replaced the encrypted detail coefficients by approximated values depending on the class they belong to. As it can be seen, even though the approximation sub-band is unknown, it is possible to identify that the encrypted image corresponds to the face of a person. There is thus a need to secure the histogram itself. The second weakness is in relation with the comparison algorithm depicted in Section \ref{ssec:2-2}. If the pirate has an idea about the dynamic bounds of $C_u^d$, and if these bounds are part of the encrypted class centers, then he will be able to determine the real value of these centers. We discuss how to address these issues in the next section.

\subsection{Encrypted wavelet coefficient histogram}\label{ssec:3-4}
The above solution allows constructing a $K$ class histogram $H_{C_u^d}$ in a clear form from an encrypted wavelet sub-band $C_u^d$. The approach we propose now aims at building a secure histogram $H_{C_u^d}^E$ the class cardinalities of which are in an encrypted form (i.e. $\{E[H_{C_d^u} (k),r_k]\}_{1\le k \le K}$) and the class centers of which cannot be ordered. It is based on two main steps. The first one consists in building a noisy histogram $H_{C_u^d}^N$ from the encrypted image wavelet coefficients; histogram from which the secure histogram $H_{C_u^d}^E$ will be computed, in a second time.

\begin{figure}[t]
\begin{center}
\includegraphics[scale=.4]{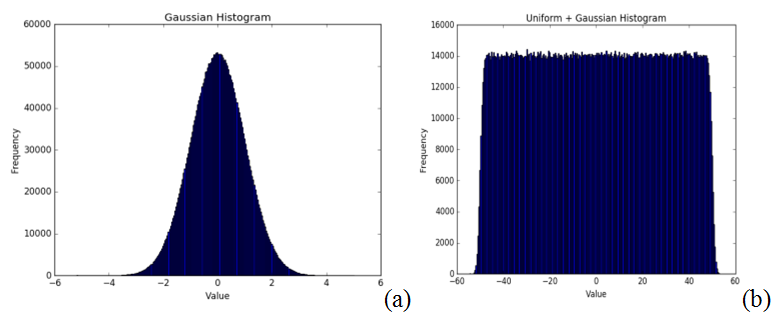}
\end{center}
\caption{\label{Fig6} Example of a noisy histogram - (a) Histogram of a discrete Gaussian random variable $X$ of mean $\nu=0$ and standard deviation $\sigma = 10$;  (b) Histogram of the random variable $Y=X+N$, where $N$ is a uniformly distributed noise in the range $[-50,50]$. }
\end{figure}

The noisy histogram construction has as purpose to secretly break the distribution of the wavelet coefficients onto a larger dynamic, while giving access to an histogram the class cardinalities of which are in a clear form (i.e., unencrypted). Our solution is equivalent to the addition of a noise to the wavelet sub-band coefficients and to compute the histogram of the resulting random variable as above. Let us consider one wavelet coefficient $C_u^d (x,y)$ and $N(x,y)$ a uniformly distributed integer in the range $[N_{min}  ,N_{max}]$ such as  $N_{max} > C_{max}$, where $C_{max}$ is the absolute maximum value a wavelet coefficient can take in a subband. The addition of $N(x,y)$ to $C_u^d (x,y)$ leads to a random variable $C_u^{dN} (x,y)=C_u^d (x,y)+N(x,y)$, the distribution of which is quite closed to a uniformly distributed random variable. In Fig. ~\ref{Fig6}, we illustrate the result of such a procedure in the case of a centered Gaussian random variable $X$ of  standard deviation $\sigma=10$ with a uniform random variable $N$ in the range $[a,b] =[-50, 50]$. The density probability $f_Y$ of the resulting random variable $Y=X+N$ takes the form 
\begin{equation}
        f_Y(t) = \frac{1}{(b-a)\sqrt{2\pi \sigma^2 }} \int_{a}^{b} \exp(-\frac{(x - t)^2}{2\sigma^2})dx
\end{equation}

From this stand point it is possible to compute the histogram of the coefficients $C_u^{dN} (x,y)$ in the encrypted domain as in the previous subsection. To do so and assuming that the dynamic of $C_u^{dN}$ is subdivided into $K'$ intervals, the user sends along with his encrypted image a set of $K'$ encrypted class centers, one set per wavelet coefficient (see Section \ref{sssec:3-3-2}). In our approach and  in order not modifying the encrypted image, which could be used for other purposes than  CBIR, it is asked to the user to add $N(x,y)$ to the histogram class centers instead of the wavelet coefficient. Adding the noise to the coefficients or to the class centers is equivalent. More clearly, for one coefficient $C_u^d (x,y)$, the user sends the set of encrypted class centers $\{E[T_k+N(x,y),r_d^u (x,y)]\}_{1\le k \le K'}$. $H_{C_u^d}^N$ will be constructed by the server using these sets of encrypted class centers. Notice, that because  $H_{C_u^d}^N$ is a uniform histogram, it gives no clues to the server about the histogram of $C_u^d$ as illustrated in the example given in Fig. \ref{(6)}. The next step is how to derive the secure histogram $H_{C_u^d}^E$, the number of class of which is $K$, from this noisy histogram of $K'$ classes.

\begin{figure}[t]
\begin{center}
\includegraphics[scale=.6]{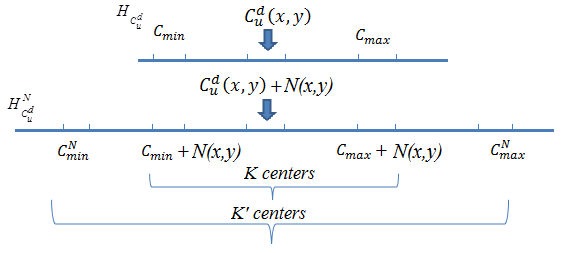}
\end{center}
\caption{\label{Fig7} Mapping between classes of $H_{C_u^d}^N$ and $H_{C_u^d}$ for a given wavelet coefficient $C_u^d (x,y)$ and a noise $N(x,y)$. The dynamic of  $C_u^{dN}$ is much greater than the one $C_u^d$ where $(K'>K)$. }
\end{figure}

The strategy we propose in order to derive $H_{C_u^d}^E$ from $H_{C_u^d}^N$ is similar to Private information retrieval (PIR) \cite{gasarch2004survey}. It  works as follows. Let us consider the computation of the cardinality of the $k^{th}$ class of the encrypted histogram $H_{C_u^d}^E (k)$. In a first time and for one coefficient $C_u^d (x,y)$, the user generates a vector $P_{x,y}^{T_k}$ of $K'$ components; components that correspond to the encryption of $0$ with different random values (i.e., $E[0,r_z]$) except its $(k+N(x,y))^{th}$ component which is fixed to ${E[1,r_u]}$. In fact, $P_{x,y}^{T_k}$  indicates for the corresponding coefficient the correct position of the $k^{th}$ class accordingly to the $K'$ classes of $H_{C_u^d}^N$ (Fig. ~\ref{Fig7} illustrates such a mapping between the histogram classes of $H_{C_u^d}^N$ and \textbf{$H_{C_u^d})$}. On its side, the server generates a vector $S_{x,y}$ the $K'$ components of  which are set to $0$ except the $l^{th}$ one which indicates the class of $H_{C_u^d}^N$
to which belongs the coefficient $C_u^{de} (x,y)$. Then, it computes the inner product between $S_{x,y}$ and $P_{x,y}^{T_k}$
\begin{equation}
    S_{x,y}.P_{x,y}^{T_k} =\left\{\begin{array}{l}
  E[1,r_u] \quad if\quad C_u^d (x,y) \in C_k \\
  E[0, r_z] \quad otherwise
\end{array}  \right.
\label{(19)}
\end{equation}
we recall that $C_k$  is the $k^{th}$ class of the histogram of $C_u^d$. More clearly, if $C_u^d (x,y)$ belongs to the $k^{th}$ class of the histogram then $S_{x,y}.P_{x,y}^{T_k}$ equals to $E[1,r_u]$. On the contrary, it equals $E[0,r_z]$. By doing so, we remove the noise added to the class centers without revealing if $C_u^d(x,y)$ belongs the $k^{th}$ class of $H_{C_u^d}$. Finally, in order to compute $H_{C_u^d}^E(k)$, i.e., the encrypted cardinality of $C_k$,  the server just has to multiply the results of the inner products  obtained  for a sub-band taking advantage of the homomorphic properties of the Paillier  cryptosystem (see ~\eqref{(7)}) that is to say
\begin{small}
\begin{equation}
    H^E_{C_u^d}(k) = \prod S_{x,y}.P_{x,y}^{T_k}
                 = \prod E[q_{x,y},r_v]
                 = E[\sum q_{x,y}, \prod r_v]
   \label{(20)}
\end{equation}
\end{small}
where $q_{x,y}\in \{0,1\}$.
 \subsection{Comparison of two encrypted wavelet coefficient histograms}\label{ssec:3-5}
As exposed in Section \ref{sssec:3-3-2}, the comparison of two images $I^{(1)}$ and $I^{(2)}$ stored into the cloud in their encrypted form by two users $U_1$ and $U_2$, respectively, stands on the computation of the $L^1$-distance between the image wavelet sub-band histograms (see ~\eqref{(18)}).

Let us thus consider two encrypted histograms $H_{C_u^{d(1)}}^E$ and $H_{C_u^{d(2)}}^E$, extracted from the encrypted version of $I^{(1)}$ and $I^{(2)}$, respectively. At this stand point, it is important to notice that $H_{C_u^{d(1)}}^E$ and $H_{C_u^{d(2)}}^E$ correspond to the cardinalities $H_{C_u^{d(1)}}$ and $H_{C_u^{d(2)}}$ encrypted with different random values and different public keys. It is however possible to compute $D^e(H_{C_u^{d(1)}}^E,H_{C_u^{d(2)}}^E)=D(H_{C_u^{d(1)}},H_{C_u^{d(2)}})$ using the second strategy we proposed in Section \ref{ssec:2-2}.

In order to measure the cardinality difference between two classes, i.e., $H_{C_u^{d(1)}}^E (k)$ and $H_{C_u^{d(2}}^E (k)$, this solution requires the comparison of these two quantities with a reference value $p_k$; value $U_1$ and $U_2$ \textit{a priori} agreed on;  that should be encrypted with the same random value as in $H_{C_u^{d(1)}}^E(k)=E[H_{C_u^{d(1)}}(k),r_k^1]$ and $H_{C_u^{d(2)}}^E(k)=E[H_{C_u^{d(2)}}(k),r_k^2]$. The problem here is that $r_k^1$ and $r_k^2$ result from multiple operations involved in the computation of $H_{C_u^{d(1)}}^E (k)$ and $H_{C_u^{d(2)}}^E (k)$, respectively (see Section \ref{ssec:3-4}). In order to make this step possible without introducing interactions between the server and the users, these ones have to send some further pieces of information they ask the server to process along with the computation of the secure histogram procedure so as to get $E[p_k,r_k^1]$ and $E[p_k,r_k^2]$.
To do so, for one coefficient $C_u^d (x,y)$, both users generate a vector $P_{x,y}$ the $K'$ component of which correspond to the value $p_k$ encrypted with the same random values as for $P_{x,y}^{T_k }$ (i.e., the vector that allows the mapping in between the classes of the noisy histogram and the clear histogram - see Section \ref{ssec:2-2}). During the computation of $H_{C_u^{d(1)}}^E (k)$, for instance, the server is asked for each coefficient to compute along with ~\eqref{(19)} the inner product
\begin{equation}
    S_{x,y}.P_{x,y} =\left\{\begin{array}{l}
  E[p_k,r_u] \quad if\quad C_u^d (x,y) \in C_k \\
  E[p_k, r_z] \quad otherwise
\end{array}  \right.
\label{(21)}
\end{equation}
Then and as for computing $H_{C_u^d}^E(k)$, it has to multiply the inner products' results in order to get
\begin{small}
\begin{equation}
    E[\sum p_k, r_k^1] = \prod S_{x,y}.P_{x,y}
                 = \prod E[p_k,r_u]
                 = E[\sum p_k, \prod r_u]
   \label{(22)}
\end{equation}
\end{small}
Due to the fact the same random values have been used either for computing $H_{C_u^{d(1)}}^E (k)$ and $E[\sum p_k,r_k^1]$, we have $r_k^1=\prod r_u$.
Following the same procedure with the image of the second user, the server will get access to $E[\sum p_k,r_k^2 ]$. Assuming the images are of same dimensions, the $L^1$-distance in-between two classes is
\begin{equation}
\begin{array}{l}
D^e(H^E_{C_u^{d(1)}}(k), H^E_{C_u^{d(2)}}(k)) = D(H_{C_u^{d(1)}}(k),\sum p_k)-\\
\;\;\;\; \; \; \;\;\;\;\;\;\;\;\; \;\;\;\;\; \;\;\;\;\;\;\;\;\;\; \;\;\;\;\;\;\;\;\;\;\;\;\;\;\;D(H_{C_u^{d(2)}}(k),\sum p_k) \\
        \;\;\;\;\;\;\;\;\;\;\;\;\;\;\;\;\;\;\;\;\;\;\;\;\;\;\;\;\;\;
\;\;\;\;\;\;\;\;\;\;\;= H_{C_u^{d(1)}}(k)-H_{C_u^{d(2)}}(k)
\end{array}
\label{(23)}
\end{equation}
and the $L^1$ distance between the two encrypted histograms
\begin{equation}
D^e(H_{C_u^{d(1)} }^E,H_{C_u^{d(2)} }^E)=\sum D^e(H_{C_u^{d(1)} }^E (k),H_{C_u^{d(2)} }^E (k))
\label{(24)}
\end{equation}

\subsection{Summary of the procedure}\label{ssec:3-6}
To sum-up, our Secured Outsourced Content Based Image Retrieval requires that one user sends with his encrypted image the following different pieces of information:

\begin{itemize}
\item In order the server computes the noisy histograms $H_{C_u^d}^N$ for each DWT sub-bands, the user needs to provide for each coefficient, $K'$ Paillier encrypted class-centers.
\item In order the server  calculates the encrypted cardinality of the $k^{th}$ class of the encrypted histogram  $H_{C_u^d}^E (k)$, and to compare it with the one of another image, the user has to upload for each coefficient

\begin{itemize}
\item   One vector $P_{x,y}^{T_k}$ of $K'$ components that maps the $k^{th}$ class of the histogram  in the clear domain with the classes of the noisy histogram; $P_{x,y}^{T_k}$ will be exploited by the server so as to calculate $H_{C_u^d}^E (k)$
\item one vector $P_{x,y}$ that contains a reference value the different users agreed on and which is encrypted $K'$ times with the same random values as for $P_{x,y}^{T_k}$, vector the server will use in order to compute the $L^1$ distance in between encrypted histograms of different images.
\end{itemize}
\end{itemize}
Once all data have been received, the server can compute in a differed time the different encrypted histograms with no-needs to interact with a trusted third party or the user.

To conclude, in the case of a dyadic separable wavelet transform and images of even dimensions $n.m$, as well as a wavelet transform up to $d$ decomposition levels while considering that a sub-band histogram in the clear domain (i.e., $H_{C_u^d}$) is constituted of $K$ classes and that the noisy histogram stands on $K'$ classes, the number of Paillier encrypted integer values that the user has to sends along with the image is
\begin{equation}
    m.n.K'+2.K'.K[3\sum_{i=1}^d \frac{m}{2^i}\frac{n}{2^i} + \frac{m}{2^d}\frac{n}{2^d}]=m.n.K'.(2K+1)
 \label{(25)}
\end{equation}

\section{Discussion}\label{SectionDISC}

\subsection{Scheme security}\label{ssec:4-1}

\begin{figure}[t]
\begin{center}
\subfloat[]{\label{fig:edge-a}\includegraphics[scale=0.6]{clear_face.png}}
  \hspace{5pt}
  \subfloat[]{\label{fig:edge-a}\includegraphics[scale=0.6]{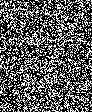}}
  \hspace{5pt}
  \subfloat[]{\label{fig:contour-b}\includegraphics[scale=0.6]{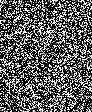}}
  \hspace{5pt}
  \subfloat[]{\label{fig:contour-c}\includegraphics[scale=0.6]{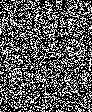}}
\end{center}
\caption{\label{Figvisage} Reconstruction attack example of the image (a) from its encrypted wavelet coefficients considering two decomposition levels and different quantization steps: (b) $\Delta=32$, (c) $\Delta=64$, (d) $\Delta=128$.}
\end{figure}

The security of our SCOBIR approach relies on the construction of secure wavelet sub-band histograms, the cardinalities of which are encrypted (i.e., $H^E_{C_u^d}= \{E[H_{C_u^d(k),r_k}] \}_{1\le k\le K}$), from a noisy histogram. Such a  noisy histogram $H^N _{C_u^d}$ is achieved by adding a uniformly distributed noise to the histogram class centers;  noise the dynamic of which is much larger than the one of wavelet coefficients(see Section \ref{ssec:3-4}). Even if $H_{C_u^d}^N$ is built by the server, i.e., this one knows to which class of $H_{C_u^d}^N$ belongs one encrypted coefficient, it has no means to identify the real class of $C_u^d$. Indeed, it does not know the class center values of $H_{C_u^d}^N$, which are encrypted, and has no idea of the mapping in between the $K'$ histogram classes of  $H_{C_u^d}^N$ with the $K$ ones of the clear histogram (i.e., $H_{C_u^d}$ ). The probability the server guess the correct class of a coefficient $C_u^d (x,y)$ is   $\frac{1}{2K}$. In fact, adding the noise to the class centers just shifts the $K$ classes of $H_{C_u^d}$  over the $K'$ ones of $H_{C_u^d}^N$. Knowing the class of $ C_u^{de}$ in $H_{C_u^d}^N$ only informs the server that $H_{C_u^d}$  has been shifted around this position for this coefficient, but this one is not able to exactly identify the class of $C_u^d (x,y)$. We give in Fig. \ref{Figvisage} the result of the reconstruction attack depicted in Section \ref{sssec:3-3-2}, where a pirate replaces encrypted coefficients that are in the same histogram class by the same arbitrary value. As shown, the image remains unintelligible. Even if the server has some \textit{a priori} knowledge about the distribution of the wavelet coefficient sub-bands, it cannot deduce any information from the noisy  histogram.

Regarding the secure wavelet sub-band histogram $H_{C_u^d}^E$, due to the fact the cardinalities of its classes are Paillier encrypted, the server has no clues about the class center ordering. Notice also that, based on the way $H_{C_u^d}^E$ is derived from $H_{C_u^d}^N$ (see Section \ref{ssec:3-4}),  the server is not able to identify the $H_{C_u^d}^E$ class of an encrypted coefficient $ C_u^{de}(x,y)$.

\subsection{Complexity and memory needs}\label{ssec:4-2}

As stated in Section \ref{ssec:3-6}, in the case of a dyadic wavelet transform and an image of even  dimensions $n.m$, the user has to encrypt  $n.m+n.m.K'.(2K+1)$ integer values (the image and the ancillary data) and send them to the server. These data will be stored in the server database. The associated computation complexity is bounded by $O(m.n.K'.(2K+1))$.
 This complexity is important but data are sent once. Notice also that all these pieces of information can be used for other purposes than SOCBIR or for computing SOCBIR with different wavelet transforms, for example.

On the server side, the computation complexity stands on the wavelet transform of the images in the encrypted domain, the construction of the different histograms and on the $L^1$-distance computation between signatures. The computation complexity of the two latter corresponds in fact to the number of comparisons the server has to do which is bounded by $O(m.n.log_2(K'))$ along with the number of modular multiplications between the results of the histogram inner-products that is bounded by $O(m.n.K)$. The server need also to perform $O(m.n.K)$ multiplications to compute the cardinality of one class of the encrypted histogram. To compute the $L^1$-distance in-between signatures, $O(LK)$ substractions are required, where $L$ is the size of database. Regarding the computation of the wavelet transform, this one imposes to conduct $m.n.log_2(K')+2.m.n.K$ multiplications and additions. As a consequence the computation complexity at the server side is bounded by $O(m.n.log_2(K')+2.m.n.K)$ modular multiplications.

Compared to the scheme presented in \cite{bellafqira2016end} which allows extracting a secure global signature from a Paillier encrypted image but with the help of two cloud providers, the complexity of which in terms of computation is of $O(m.n)$ encryption for the user and $O(m.n.K+L.K+L)$ encryptions for both servers, our approach is of $K'(2K+1)$ times greater in terms of computation complexity for the user and $log_2(K')+K$ times greater for the server. However our proposal uses only one server and is free of communication rounds while \cite{bellafqira2016end} require $3$ communication rounds (i.e., data transmission, signature computation and signature comparison).

  \section{Experimental Results}\label{Section4}
  \subsection{Test databases and performance criteria}\label{ssec:5-1}

Our SOCBIR scheme has been experimented considering two
application frameworks: diagnosis aid support and authentication based on
face recognition (see Section \ref{intro}). In both cases, data are  stored
encrypted in the cloud. For a new image (a medical image to
interpret or a user to recognize), the user system encrypts it and
sends it to the server along with all pieces of information
depicted in Section \ref{ssec:3-6}. The server will send back the most
similar images it will have identified in its database.

\begin{figure}[t]
\begin{center}
  \subfloat[]{\label{fig:edge-a}\includegraphics[scale=1.5]{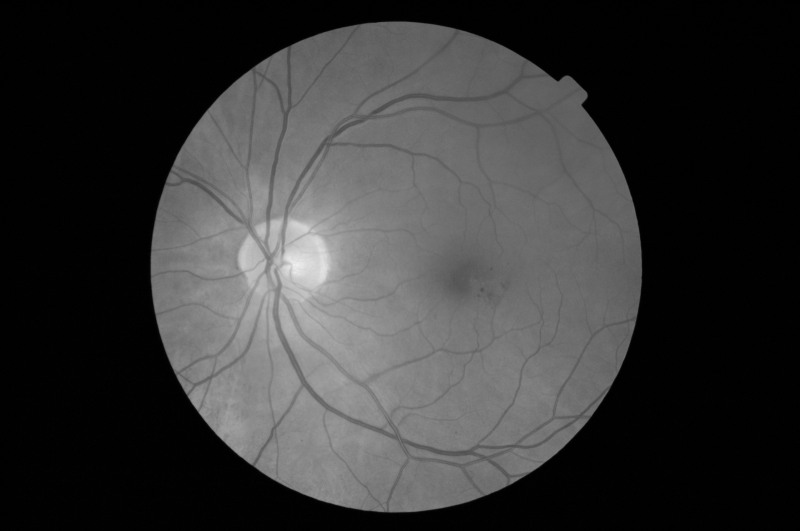}}
\hspace{5pt}
  \subfloat[]{\label{fig:edge-a}\includegraphics[scale=.35]{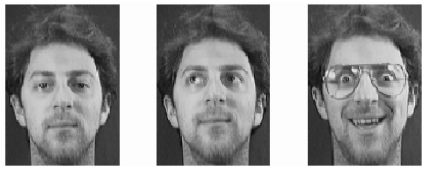}}
  \hspace{5pt}
  \end{center}
\caption{\label{Fig8}Illustrative samples of our image test sets (a) retina image, (b) facial images of one user. }
\end{figure}

We give in Fig. \ref{Fig8} some image samples issued from our two
test image databases. The medical database regroups $1200$ $8$ bit
depth retina images of  $2240\times 1488$ pixels associated with
pieces of information about the lesions present into each images.
The face recognition database is issued from the Olivetti Research
Laboratory of Cambridge, UK. It contains $400$ images of $92\times
122$ pixels  that correspond to $10$ photographs of $40$ distinct
subjects.

In both use-cases, the performance of our SOCBIR scheme is evaluated in terms of CBIR mean precision which corresponds to the rate of returned images with the same pathology or with the same person as in the query image. As we will see, this measure will vary upon the number of classes considered into the histograms.


\begin{figure}[t]
\begin{center}
  \subfloat[]{\label{fig:edge-a}\includegraphics[scale=2.25]{retina.png}}
  \hspace{5pt}
  \subfloat[]{\label{fig:edge-a}\includegraphics[scale=2.25]{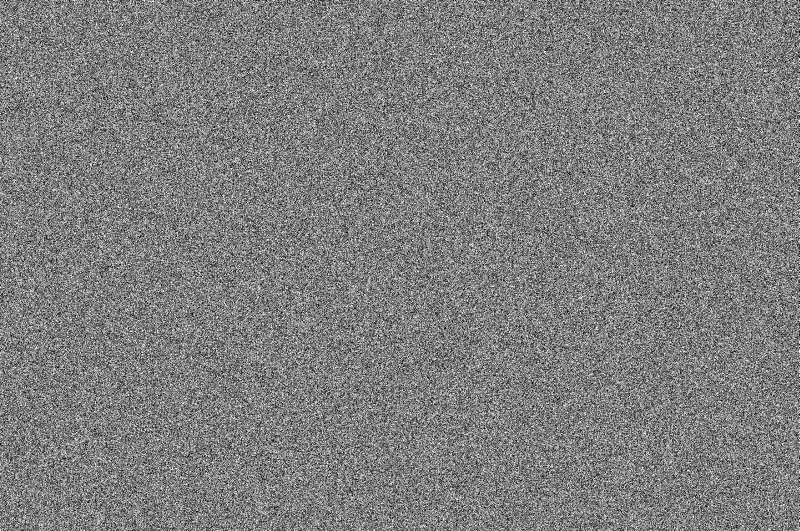}}
\hspace{5pt}
  \subfloat[]{\label{fig:contour-c}\includegraphics[scale=0.28]{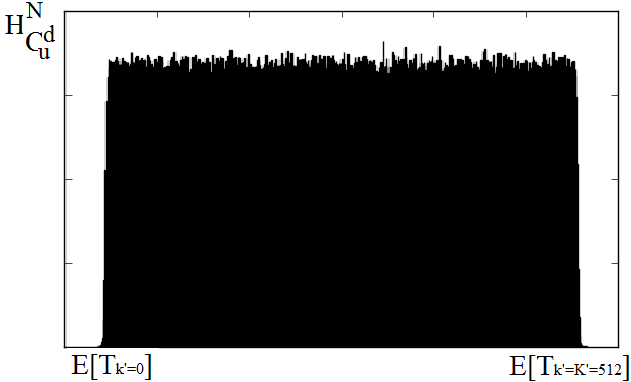}}
\hspace{5pt}
  \subfloat[]{\label{fig:contour-b}\includegraphics[scale=0.28]{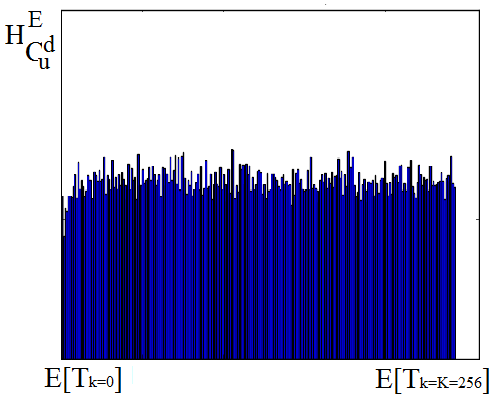}}
\end{center}
\caption{\label{Fig9} Encrypted version (b) of the retina image (a). (c) noisy histogram of the vertical detail sub-band at the first image decomposition level (i.e., of the coefficients $C^{d=1}_{HG}$) with a uniform random noise in the range $[-128,128]$, a quantization step $\Delta=1$, considering $K'=512$ classes. (d) Encrypted histogram of the same sub-band with $K=256$ classes.}
\end{figure}

\subsection{SOCBIR scheme performance with medical images}\label{ssec:5-2}

    In this experiment, the $2D$ Haar wavelet transform (HWT) has been considered. In order to make its computation possible in the encrypted domain up to the $2^{nd}$ resolution level (i.e. $d= 0, 1, 2 $), its decomposition filters' coefficients have been turned into integer values by fixing the quantization factor $Q$ of ~\eqref{(12)} to $4$ (see Section \ref{ssec:3-2}). The signature of one image corresponds to the sub-band histograms up to a given resolution level, including the histogram of the approximation sub-band. The random variable added to the class center values is in the range $[-128,128]$.

Before detailling the retrieval performance of our scheme, let us first come back on the amount of data the user has to send. In the case images are Paillier encrypted with large prime numbers $p$ and $q$ such that their product, i.e., the public key $K_p$ (see \eqref{(1)}), stands on 1024 bits, then one encrypted integer will be 2048 bit encoded. A direct consequence of such memory needs is that it is difficult to conduct large experimentations with such public key size on a single computer. In order to reduce these needs, all the following tests have been conducted with a public key of 16 bits. The retrieval performance of our scheme with a key of 1024 bits can be directly derived from these results as they are independent of the public key size. Results will be the same.
We give in Fig. \ref{Fig9}, a view of an encrypted retina image accompanied of the noisy and encrypted histograms of its vertical detail sub-band of its first wavelet decomposition level. It can be seen that these histograms are uniformly distributed as expected.

\begin{figure}[t!]
\begin{center}
\includegraphics[scale=.6]{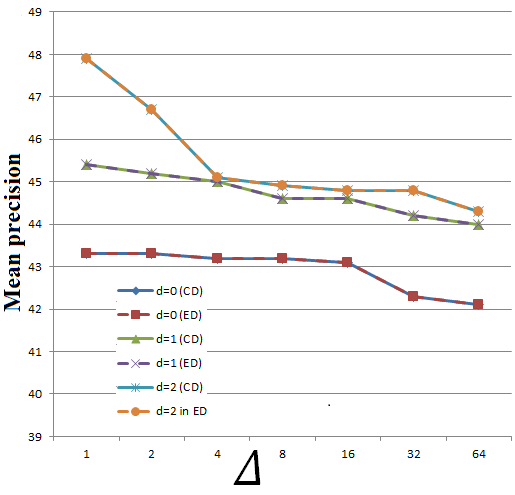}
\end{center}
\caption{\label{Fig11}Retrieval performance expressed in mean precision at five (in $\%$) for  our medical database considering different quantization step values ($\Delta$) and distinct decomposition level $(d)$. Dashed curves denote SOCBIR results in the encrypted domain (ED) while full curves give the performance of the equivalent CBIR system in the clear domain (CD). }
\end{figure}

Fig. \ref{Fig11} provides the retrieval performance of our scheme for a mean precision at five, that is to say that, for one request image, the server returns the five most similar images it finds in its database. We also compare our scheme with the same CBIR approach in the clear domain considering different wavelet decomposition levels (i.e., $d=\{0,1,2\}$) and several histogram quantization step values (i.e.,  $\Delta \in [1, 2, 4, 8, 16, 32, 64]$). Notice also that all curves are plotted in average, using $200$ images of our retina image test set as user request images; the 800 other images constitute the server database. As it can be seen our SOCBIR scheme has the same performance than the CBIR scheme in the clear domain. Such a result can be explained based on the fact the filter coefficients of the Haar wavelet transform in the encrypted and the clear domains are equivalent. More clearly, the coefficient expansion \eqref{(12)} does not impact the wavelet transform precision. This may not be the case with wavelet transforms the filter coefficients of which are real numbers. Indeed, a loss of precision in the wavelet coefficient computation could reduce the performance retrieval. Beyond, if in the clear domain one may work with $\Delta=1$, i.e., working with the most accurate histogram based signature, using values of $\Delta$ at least $16$ times greater do not impact so much retrieval performance. Let us also recall that the smallest $\Delta$, the more precise is the histogram and the greater is the number of encrypted histogram class centers the user has to send along with his image.

\subsection{SCOBIR performance for face recognition}\label{ssec:5-3}

\begin{figure}[t]
\begin{center}
\includegraphics[scale=.5]{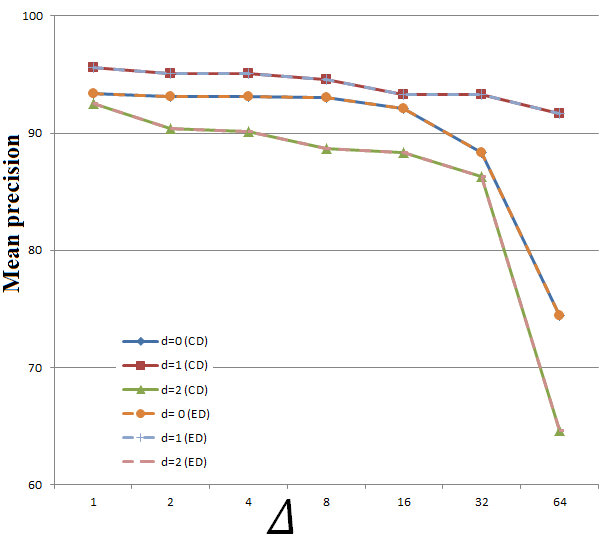}
\end{center}
\caption{\label{Fig12} Retrieval performance expressed in mean precision at five (in $\%$) for our face database considering  different  quantization step values ($\Delta$) and distinct decomposition level $(d)$. Dashed curves denote SOCBIR results in the encrypted domain (ED) while full curves give the performance of the equivalent CBIR system in the clear domain (CD). }
\end{figure}

\begin{figure}[h]
\begin{center}
\includegraphics[scale=.35]{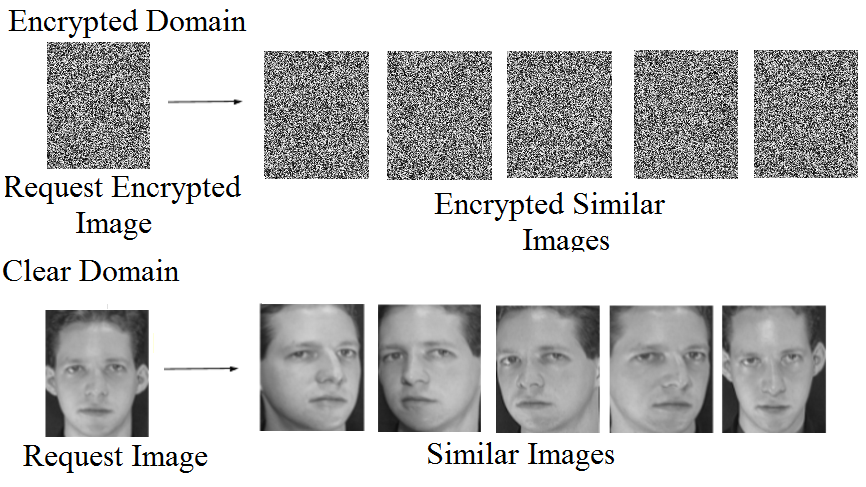}
\end{center}
\caption{\label{fig14} Illustrative example of the response of our SOCBIR scheme to a request for user authentication.}
\end{figure}

This experiment has been conducted with the same parametrization as the previous one. Again, our test image database has been split into two sets: 200 facial images are used as request image set while the 200 others are used as reference images by the server. The mean precision at five of our scheme and the equivalent approach in the clear domain is given in Fig. \ref{Fig12}, in average, depending on different quantization step values (i.e.,  $\Delta \in [1, 2, 4, 8, 16, 32, 64]$) and decomposition levels (i.e., $d=\{0,1,2\}$). The response of our scheme to one request image is also illustrated in Fig. \ref{fig14}. Again, one can see that our SOCBIR approach achieves the same performance as the CBIR approach in the clear domain whatever the quantization step and the level of decomposition. Based on the fact facial images are much smaller than retina images, the amount of information the user has to send to the server is reduced. This is the same for the time computation when comparing images.

\section{Conclusion}\label{SectionCCL}
In this paper, we have proposed a novel Secure Outsourced  CBIR method, which allows completely carrying out a search in an encrypted image database maintained by a server. Contrarily to actual homomorphic encryption based CBIR schemes, it allows the extraction of an encrypted global image signature and does not require extra communications between the user and the server or a trusted third party during the search computation. It exploits a secured histogram procedure we proposed so as to build encrypted wavelet coefficient histograms from an encrypted image taking advantage of a fast comparison of Paillier encrypted data and a noisy histogram. Experimental results show that our SOCBIR scheme achieves the same performance as the equivalent CBIR approach in the clear domain. If our solution guarantees data confidentiality and good retrieval performance with zero communication, its use in real practice remains limited due to the fact it requires the user to send Gigabits of encrypted data when using a public key of 1024 bits so as to be secure. Future works will focus on reducing this amount of information.
\section*{Acknowledgment}

This work has been supported by the French national research agency through the LabCom SePEMeD, ANR-13-LAB2-0006-01, and by the Brittany Region Council, ARED AR-MEM.

\ifCLASSOPTIONcaptionsoff
  \newpage
\fi



\bibliographystyle{IEEEtran}

%

\bibliographystyle{ieeetran}
\bibliography{SCBIR}


%








\end{document}